\documentclass{moriond}

\usepackage{comment}
\def\be{\begin{equation}}
\def\ee{\end{equation}}
\def\bea{\begin{eqnarray}}
\def\eea{\end{eqnarray}}

\newcommand{\Photo}{\includegraphics[height=80mm]{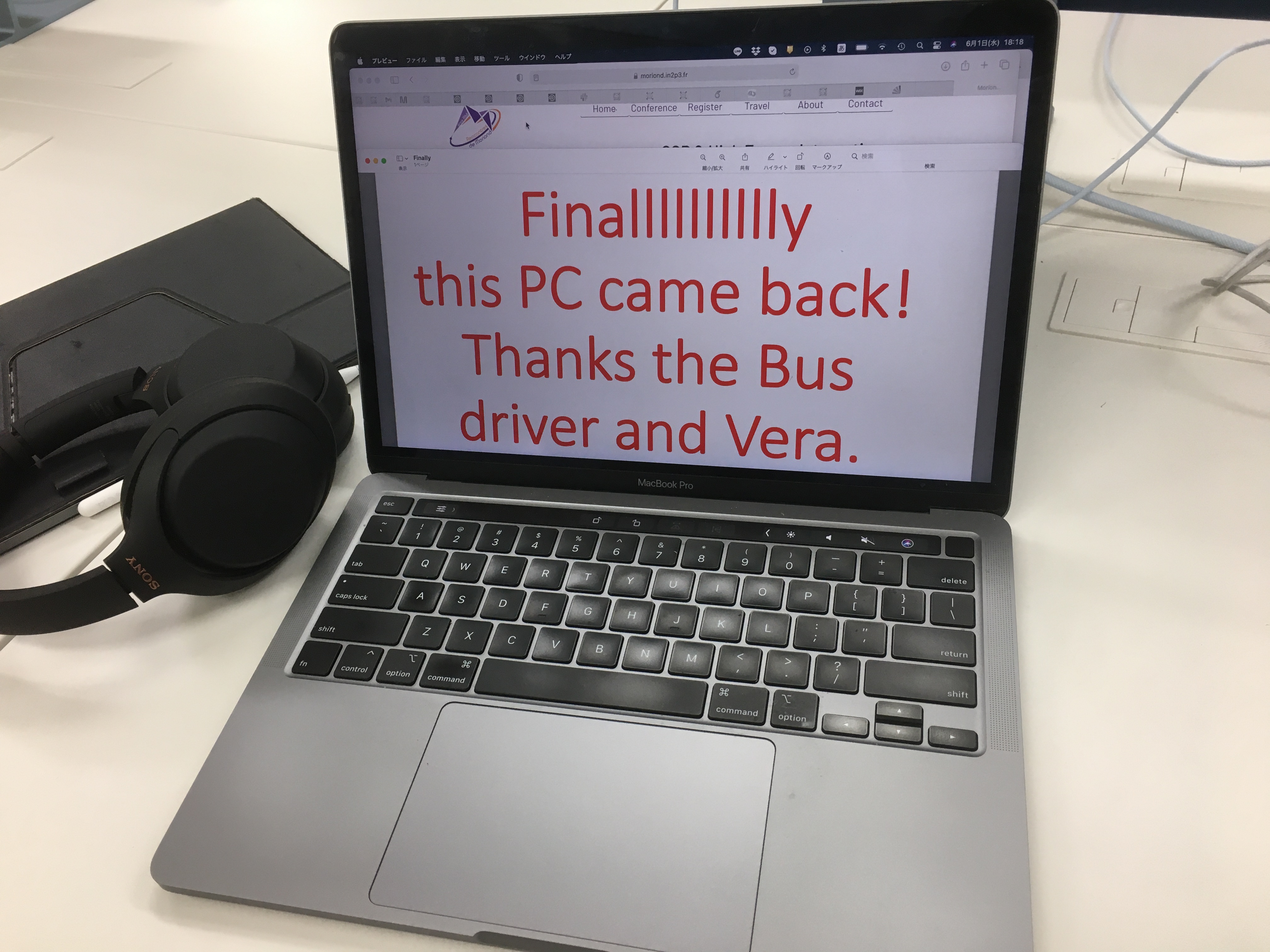}}

\begin{document}
\begin{flushright}
 {\tt
TTP22-038, P3H-22-061
}
\end{flushright}

\vspace*{2cm}
\title{Interplay between the $R_{D^{(*)}}$ anomaly and the LHC }

\author{ Syuhei Iguro }

\address{Institute for Theoretical Particle Physics (TTP), Karlsruhe Institute of Technology (KIT),
Engesserstra{\ss}e 7, 76131 Karlsruhe, Germany,\\ Institute for Astroparticle Physics (IAP),
Karlsruhe Institute of Technology (KIT), 
Hermann-von-Helmholtz-Platz 1, 76344 Eggenstein-Leopoldshafen, Germany}

%%%%%%%%%%%%%%%%%%%%%%%%%%%%%%%%%%%%%%%%%%%%%%%
\maketitle\abstracts{
The current discrepancy in $R_{D^{(*)}}=BR(\bar{B}\to D^{(*)} \tau\bar{\nu})/BR(\bar{B}\to D^{(*)} \ell\bar{\nu})$ where $\ell=e,\,\mu$ may imply a new particle with the mass around the order of TeV.
Therefore it is well motivated to test those scenarios at the large hadron collider (LHC).
Thanks to the size of the discrepancy and the analysis based on the effective field theory (EFT), the variety of the possible new physics candidates is limited.
We focus on the charged Higgs ($H^-$) and leptoquark ($X$) possibility and discuss the collider physics.  }
%%%%%%%%%%%%%%%%%%%%%%%%%%%%%%%%%%%%%%%%%%%%%%%

\section{Introduction}
The lepton flavor universality is one of the most important predictions of the standard model (SM), however, a new physics often interacts differently depending on the flavor.
In the $\bar{B}\to D^{(*)} \tau\bar{\nu}$ decay, the uncertainty in the theoretical prediction mainly stems from the Cabbibo-Kobayashi-Maskawa matrix element,
%\cite{Cabibbo:1963yz,Kobayashi:1973fv}
$V_{cb}$ and $\bar{B}\to D^{(*)}$ meson transition form factors.
%\cite{Isgur:1989vq}.
Thanks to the recent both theoretical and experimental developments, the more dedicated form factor parameterization and their determination
% beyond the conventional CLN\cite{Caprini:1997mu} and BGL\cite{Boyd:1997kz}
is now possible\cite{Bordone:2019vic,Iguro:2020cpg}. 
Since taking the ratio (almost) cancels the uncertainty in the former (latter), it is expected that $R_{D^{(*)}}$ is a good window to access the new physics.
It has been almost 10 years since the astonishing deviation was firstly reported by the BaBar experiment which is followed by the LHCb and Belle experiments.
However, recently the world average has been coming closer to the SM prediction, the significance of the deviation is still more than 3-4 $\sigma$ due to the reduced uncertainties.
The current experimental world average (WA)\cite{Aoki:2021kgd} and the SM prediction\cite{Iguro:2020cpg} are summarized as 
\begin{eqnarray}
R_{D}^{{\rm{WA(SM)}}}=0.339\pm0.030~(0.248\pm0.001),~~R_{D^*}^{{\rm{WA(SM)}}}=0.295\pm0.014~(0.289\pm0.004),
\end{eqnarray} 
and the lepton flavor universality among e and $\mu$ is confirmed at $\mathcal{O}$(1)$\%$.
%\cite{Belle:2018ezy}

It is known that the modification in the tauonic mode is necessary if the deviation is taken as a hint for the new physics.
Within the SM the tree level W boson exchange describes the $\bar{B}\to D^{(*)}\tau\bar\nu$ transition and the required size of the shift in $R_{D^{(*)}}$ is 10--20$\%$.
Because of the magnitude of the shift, we can safely focus on the operators up to dimension 6 and then the variety of the possible new mediator particles is limited. 
See Ref.\,\cite{London:2021lfn} for a recent review.
Once we assume the new particle couples to the left handed neutrino, only two categories of the scenarios are available.
One is a charged Higgs($H^-$) and the other is a leptoquark ($X$).
In this proceedings, we will discuss the LHC phenomenology of the those scenarios. 
Since the interactions to enhance the $R_{D^{(*)}}$ automatically generates $\tau\nu$ final state we mainly focus on the $\tau\nu(+b)$ signature. 
%%%%%%%%%%%%%%%%%%%%%%%%%%%%%%%%%%%%%%%
%%%%%%%%%%%%%%%%%%%%%%%%%%%%%%%%%%%%%%%
\section{Charged Higgs}
%%%%%%%%%%%%%%%%%%%%%%%%%%%%%%%%%%%%
\begin{figure}[t]
\begin{minipage}{0.50\linewidth}
\centerline{\includegraphics[width=1\linewidth]{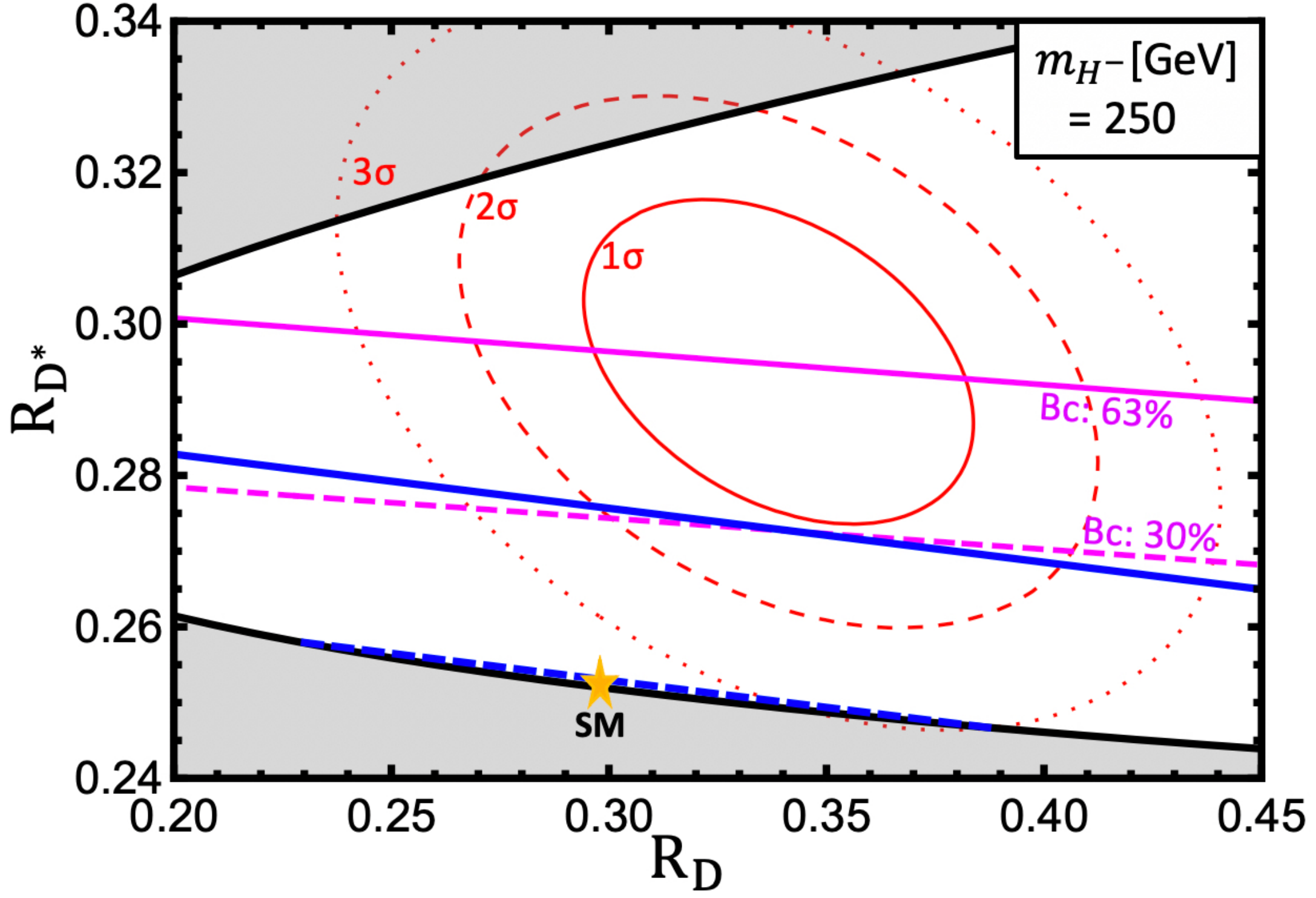}}
\end{minipage}
\begin{minipage}{0.50\linewidth}
\centerline{\includegraphics[width=1\linewidth]{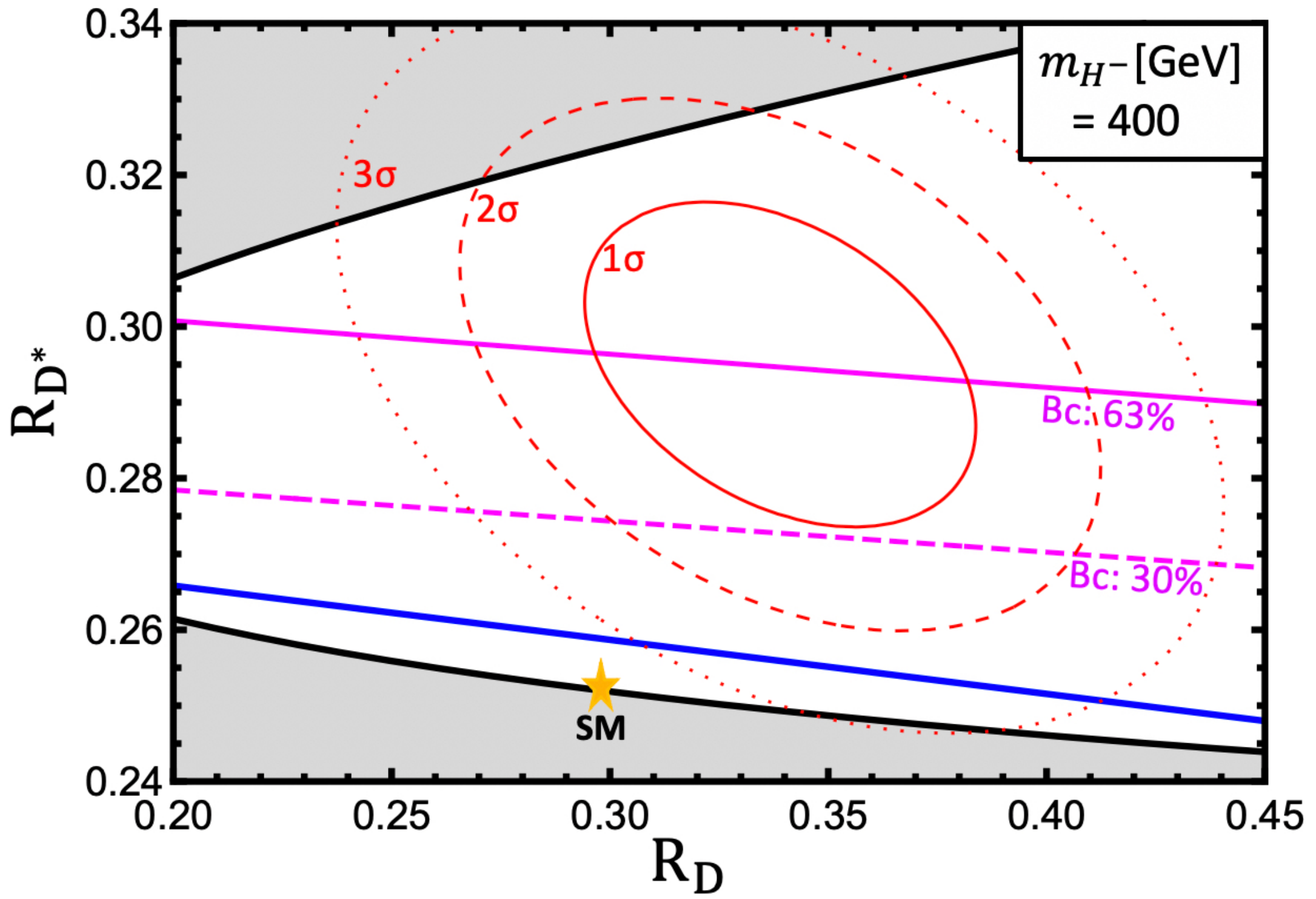}}
\end{minipage}
\caption{The impact of the $\tau\nu+b$ search on the charged Higgs with $m_{H^-}=250$\,GeV(left) and $m_{H^-}=400$\,GeV(right).
See the main text for the detail.}
\label{fig:result}
\end{figure}
%%%%%%%%%%%%%%%%%%%%%%%%%%%%%%%%%%%%
It is well known that the charged Higgs explanation of the discrepancy easily enhances the decay width of the $B_c^-\to\tau\bar{\nu}$ decay\cite{Celis:2012dk,Iguro:2017ysu,Iguro:2018qzf}.
The upper limit on BR$(B_c^-\to\tau\bar{\nu})\le 30(10)\%$ is derived from the lifetime of the $B_c$ meson\cite{Alonso:2016oyd} and the LEP data\cite{Akeroyd:2017mhr}, respectively. 
However, it was pointed out that the large charm mass uncertainty and energy dependence of the fragmentation function relax the bound as large as BR$(B_c^-\to\tau\bar{\nu})\le 60\%$\cite{Blanke:2018yud}.
Recently the reevaluation in Ref.\cite{Aebischer:2021ilm} found BR$(B_c^-\to\tau\bar{\nu})\le 63\%$ based on the $B_c$ lifetime . 

Testing the charged Higgs scenario at the LHC is easier than that for the other scenarios since the larger coupling is necessary to enhance $R_{D^*}$ compared to the other scenarios due to small coefficients.\footnote{See, Eq.\,(2.4) of Ref.\cite{Iguro:2018vqb} for instance.}
In Ref.\,\cite{Iguro:2018fni} authors reinterpreted the bound on the heavy resonance reported by the CMS to derive the constraint on the charged Higgs scenario based on the fast simulation.
It was found that the result with 36fb$^{-1}$ can already exclude the explanation with $m_{H^-}\ge\,400$\,GeV where the data is available.
In light of the relaxed bound from the $B_c$ decay, Ref.\,\cite{Iguro:2022uzz} revisited the light mass window 180\,GeV$\le m_{H^-}\le$ 400\,GeV with the $\tau\nu$ constraint based on Run 1 data, the stau constraint and low mass di-jet constraint at the LHC.
However, it turned out that this mass region is difficult to fully cover the interesting parameter region even at the high luminosity LHC.
More recently it is shown that requiring an additional b-tagged jet in a low mass $\tau\nu$ resonance search, which is not performed experimentally, can enhance the signal to back ground ratio\cite{Blanke:2022pjy}.
In a conventional $\tau\nu$ resonance search, the W boson tail consists of a dominant background.
By the requiring the additional b-jet, we can suppress the valence quark contribution.
For instance $ug\to W^*b\to \tau\nu+b$ contribution is suppressed by the factor of $|V_{ub}|^2$.
Furthermore the $j\to b$ miss-tagged contribution in $gu\to W^*j\to \tau\nu+j$ is suppressed by the small miss tagging rate of $\mathcal{O}$(10$^{-3}$).
As a result, the huge background reduction of a factor of $\mathcal{O}$(100) is observed while the signal reduction is found to be of a factor.

In fig.\,\ref{fig:result}, the sensitivity with the current luminosity and the projected sensitivity with the HL-LHC are shown in blue solid and dashed lines on the $R_D$ vs $R_{D^*}$ plane.
The charged Higgs mass is assumed to be 250 (400)\,GeV on the left (right) panel.
The area above the line can be probed and the dashed line is omitted on the right since the line almost degenerates with the SM prediction.
The world average of the $R_{D^{(*)}}$ data at 1, 2 and 3$\sigma$ are shown by the red solid, dashed and dotted contours.
The grey shaded region is out of the prediction within the model.
The SM prediction is indicated by a yellow star, and the horizontal magenta solid (dashed) line corresponds to BR($B_c\to\tau\nu)=63\,(30)\%$ as a comparison. 

The figure shows that the collider prospect with 139\,fb$^{-1}$ of the data can cover the broader range and judge the $1\,\sigma$ explanation of the anomaly within the model.
%%%%%%%%%%%%%%%%%%%%%%%%%%%%%%%%%%%%%%%
\section{Leptoquark}
%%%%%%%%%%%%%%%%%%%%%%%%%%%%%%%%%%%%
\begin{figure}[t]
\begin{minipage}{0.50\linewidth}
\centerline{\includegraphics[width=1\linewidth]{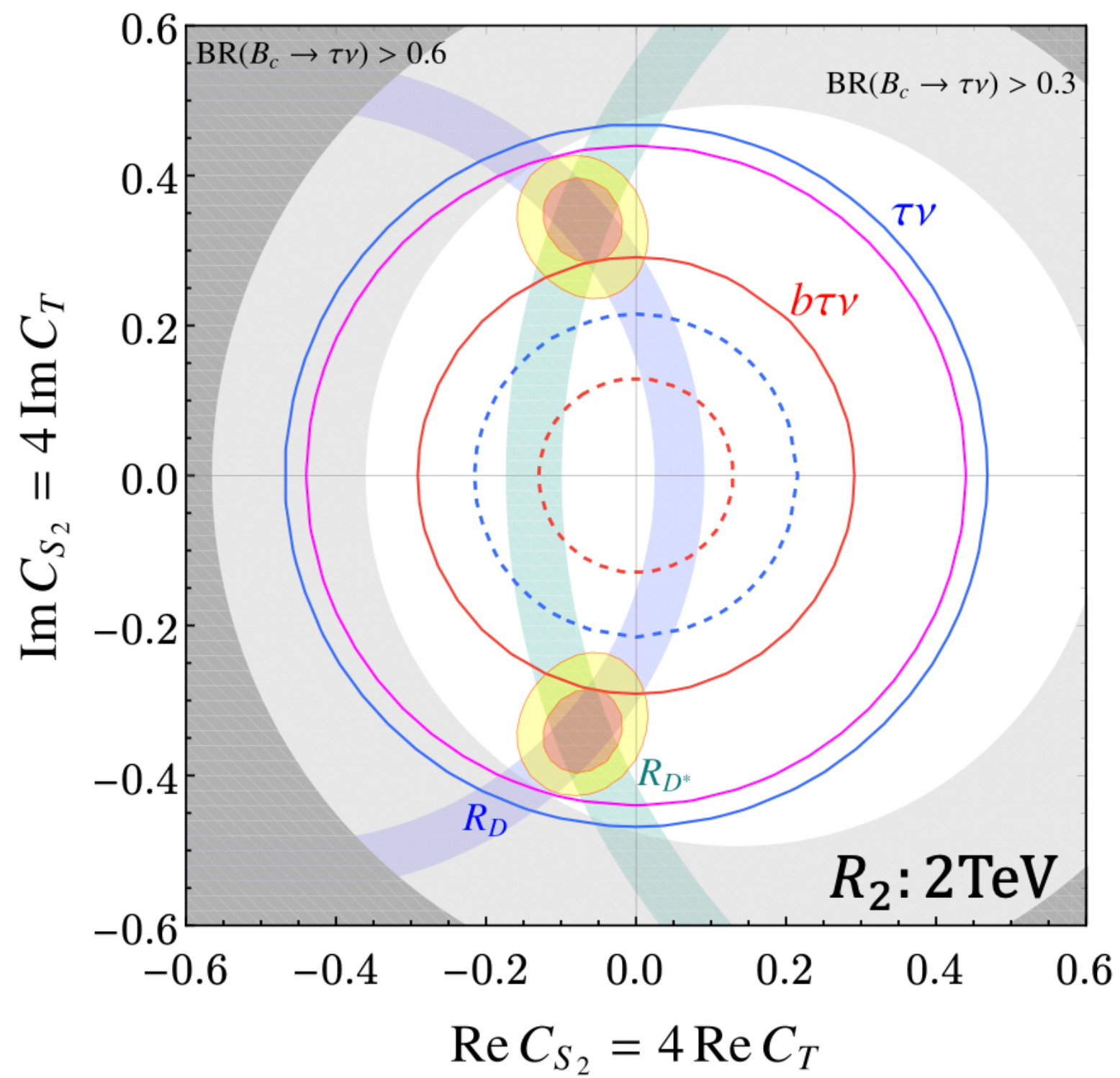}}
\end{minipage}
\begin{minipage}{0.50\linewidth}
\centerline{\includegraphics[width=1\linewidth]{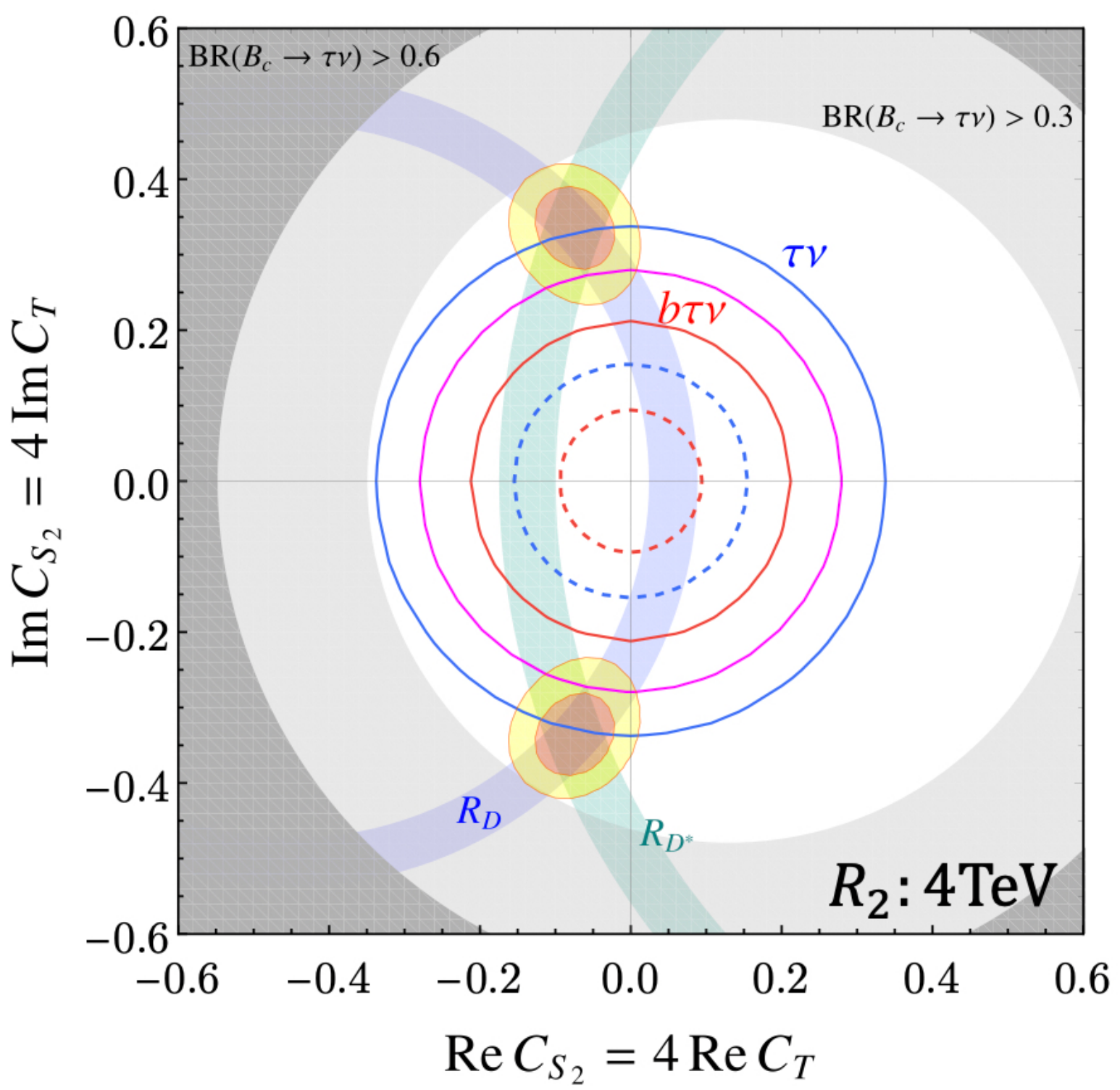}}
\end{minipage}
\caption{The impact of the $\tau\nu$ and $\tau\nu+b$ searches on the $R_2$ leptoquark scenario where $m_{R_2}=2$\,TeV(left) and $m_{R_2}=4$\,TeV(right) are assumed.
See the main text for the detail. }
\label{fig:result2}
\end{figure}
%%%%%%%%%%%%%%%%%%%%%%%%%%%%%%%%%%%%
There are three types of the leptoquarks, $X=R_2$, $S_1$ and $U_1$ which can explain the discrepancy.
The charge assignment under the SM gauge symmetry (SU(3)$_c$, SU(2)$_L$, U(1)$_Y$) is given as ($3,~ 2, ~7/6$) for $R_2$, ($\bar{3},~1,~1/3$) for $S_1$ and ($3, 1, 3/2$) for $U_1$, respectively.

Different from the charged Higgs case, a leptoquark predicts the non resonant $\tau\nu$ signature and the high $p_T$ region is sensitive to the new physics effect since the smaller background contribution is expected there.
The significant leptoquark mass dependence is pointed out even with the t-channel mediator because the large momentum exchange which corresponds to the large Mandelstem variable, $t$ is necessary to generate the high $p_T$ leptons\cite{Iguro:2020keo}.
Furthermore it is shown that selecting the negatively charged $\tau$ can improve the sensitivity.
This is because that the density of the up quark is larger than that of down quark.
Also the additional b-tagged helps to enhance the sensitivity to leptoquark scenarios by 30-40$\%$\cite{Endo:2021lhi}.
It is noted that the $\tau\bar{\tau}(+b)$ final state also provides the similar sensitivity\cite{CMS:2022rbd}.

In the following we focus on the $R_2$ leptoquark scenario as a demonstration.
The scenario predicts $C_{S_2} = +4C_T$ at the leptoquark mass scale where $C_{S_2}$ and $C_T$ are Wilson coefficients of a scalar and tensor operators which describes the $b\to c\tau\bar{\nu}$ transition.
In fig.\,\ref{fig:result2}, the constraint and prospect in $\tau\nu$ and $\tau\nu+b$ modes are shown with $m_{R_2}=2$\,TeV(left) and $m_{R_2}=4$\,TeV(right).
The regions outside the blue and red lines can be probed with the $\tau \nu$ and $\tau\nu+b$ signature, respectively, where the solid and dashed lines correspond to the prospect based on 139fb$^{-1}$ and 3ab$^{-1}$ of the data.
The magenta line shows the bound with the previous CMS result with 36fb$^{-1}$ derived in Ref.\cite{Iguro:2020keo}.
It is noted that the CMS found the smaller event number compared to their expectation, and hence it results in the stringent constraint. 
The darker and lighter gray shaded regions are constrained by BR$(B_c \to \tau \nu)\le0.6$ and BR$(B_c \to \tau \nu)\le0.3$.
The $R_D$ and $R_{D^\ast}$ anomalies are explained at $1\,\sigma$ in the blue and green shaded regions, respectively.
The combined fits at 1 and 2\,$\sigma$ are shown in orange and yellow, respectively.

It is found that the LHC sensitivity of the $\tau\nu$ search is marginal with the Run 2 full data to probe the $R_{D^{(*)}}$ explanation depending on the mass, 
while that of the $\tau^\pm\nu+b$ search is enough to probe the 1$\sigma$ parameter region
in both $m_{R_2}=2$\,TeV and $m_{R_2}=4$\,TeV cases.
Therefore, it is concluded that requiring an additional $b$-jet is significant to test a leptoquark scenario in light of the $R_{D^{(*)}}$ anomaly.
The study for the other leptoquark scenarios can be found in Refs.~\cite{Iguro:2020keo,Endo:2021lhi}.
  
%%%%%%%%%%%%%%%%%%%%%%%%%%%%%%%%%%%%%%%
\section{Conclusion}
In this proceeding we discussed the LHC phenomenology of the viable solutions to the notorious $R_{D^{(*)}}$ anomaly.
We showed that additional b-tagging in a $\tau\nu$ resonance search can improve the sensitivity to a possible new physics explanation for the discrepancy.
The remaining low mass charged Higgs window which recently revived thanks to the relaxed BR($B_c\to\tau\nu$) bound can be probed by looking for the event with $\tau\nu$ and an additional b-tagged jet.
We see that the collider probe can provide the more powerful tool for the scenario with the Run 2 data.
Also requiring an additional b-tagged jet helps to enhance the sensitivity to the leptoquark scenarios.
As a demonstration we discussed the $R_2$ leptoquark solution and found that the additional b-tagging is crucial to test the scenario with the current amount of the data.  
%%%%%%%%%%%%%%%%%%%%%%%%%%%%%%%%%%%%%%%

%%%%%%%%%%%%%%%%%%%%%%%%%%%%%%%%%%

\section*{Acknowledgments}
I would like to thank the organizers for the generous support, hospitality.
Especially I appreciate Vera for helping me to find the laptop which I left in the conference bus to Geneva station.  
This work is supported by the Japan Society for the Promotion of Science (JSPS) Core-to-Core Program, No.JPJSCCA20200002 and the Deutsche Forschungsgemeinschaft (DFG, German Research Foundation) under grant 396021762-TRR\,257.

\end{document}